\documentclass[twocolumn,showpacs,superscriptaddress,floatfix]{revtex4-1}
\usepackage{amsmath}
\usepackage{graphicx}
\usepackage{color}
\usepackage{amstext}
\usepackage{hyperref}

\begin{document}

\title{Mechanism for turbulence proliferation in subcritical flows}

\author{Anna \surname{Frishman}}
\affiliation{Technion Israel Institute of Technology, 32000 Haifa,
  Israel}
\author{Tobias \surname{Grafke}}
\affiliation{University of Warwick, Coventry CV4 7AL, United Kingdom}

\date{\today}

\begin{abstract}
  The subcritical transition to turbulence, as occurs in pipe flow, is
  believed to generically be a phase transition in the directed
  percolation universality class. At its heart is a balance between
  the decay rate and proliferation rate of localized turbulent
  structures, called puffs in pipe flow. Here we propose the
  first-ever dynamical mechanism for puff proliferation---the process
  by which a puff splits into two. In the first stage of our
  mechanism, a puff expands into a slug. In the second stage, a
  laminar gap is formed within the turbulent core. The notion of a
  split-edge state, mediating the transition from a single puff to a
  two puff state, is introduced and its form is predicted. The role of
  fluctuations in the two stages of the transition, and how splits
  could be suppressed with increasing Reynolds number, are
  discussed. Using numerical simulations, the mechanism is validated
  within the stochastic Barkley model. Concrete predictions to test
  the proposed mechanism in pipe and other wall bounded flows, and
  implications for the universality of the directed percolation
  picture, are discussed.
\end{abstract}

\maketitle

\section{Introduction}
\label{sec:background}

How pipe flow becomes turbulent, a seemingly mundane phenomenon, has
been a lasting puzzle for more than a century~\cite{reynolds:1883}. As
first recognized by Reynolds, pipes have a transitional flow regime,
where localized turbulent structures and laminar flow
coexist~\cite{lindgren:1957,wygnanski:1973,darbyshire:1995}.  However,
a clear understanding of the nature of these structures, called puffs,
and of the transition to turbulence with increasing Reynolds number
$Re$, has only emerged in the past
decade~\cite{avila:2011,barkley:2015,barkley:2016}.  The current view
is that it is an out-of equilibrium phase transition lying in the
directed percolation universality class~\cite{pomeau:1986}, and
moreover that it is the ubiquitous route to turbulence for wall
bounded flows.

For pipe flow, the extremely long time-scales and length-scales
involved prevent a direct confirmation of this
picture~\cite{mukund:2018}. It has, however, been confirmed in other
wall bounded flows---which share much of the phenomenology of pipe
flow~\cite{manneville:2016,tuckerman:2020}, and where the critical
point is more
accessible~\cite{shi:2013,lemoult:2016,chantry:2017,tuckerman:2020,klotz:2022}.

Puffs are the basic degrees of freedom in the transitional
picture. The fraction of the pipe occupied by puffs determines the
level of turbulence, which is the order parameter for the
transition. The absorbing state, required for a directed percolation
transition, is the laminar flow: turbulent puffs cannot be
spontaneously excited from it. The spatial proliferation of turbulence
can thus only occur through puff splitting---a rare and random process
by which two puffs are generated from a single puff. This process,
however, competes with random decays of puffs, returning the flow back
to the laminar state. The opposing tendencies with $Re$ of these two
processes bring about the critical point: decays become rarer with
increasing $Re$, while splits become more frequent, with the critical
point occurring roughly where the rates of the two
balance~\cite{avila:2011,barkley:2016}.

\begin{figure}
  \begin{centering}
    \includegraphics[height=130pt]{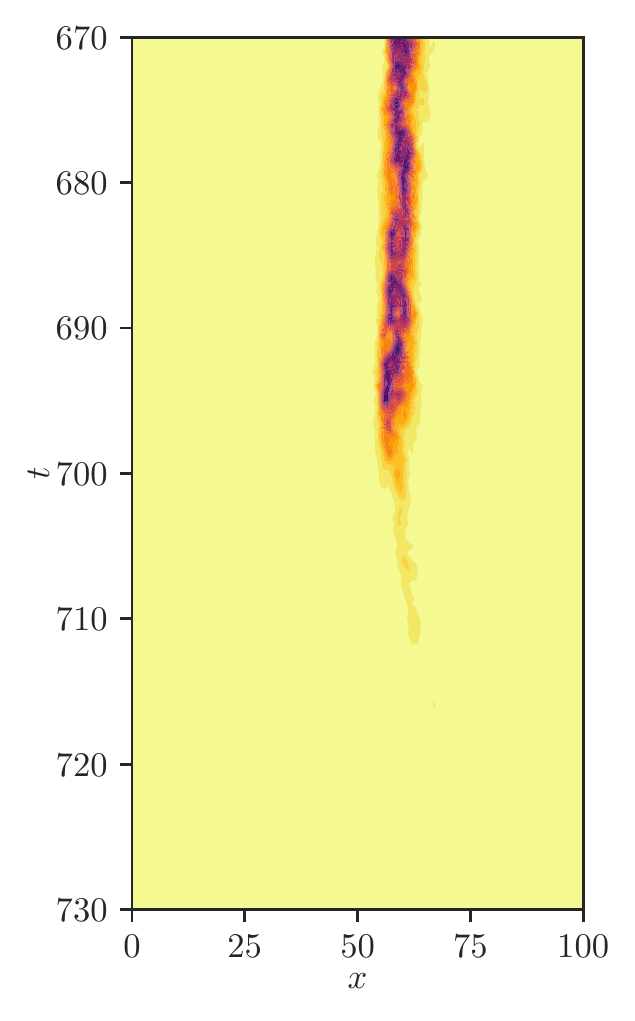}\hspace{-8pt}
    \includegraphics[height=130pt]{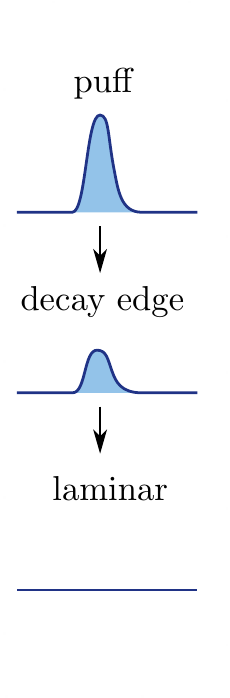}
    \includegraphics[height=130pt]{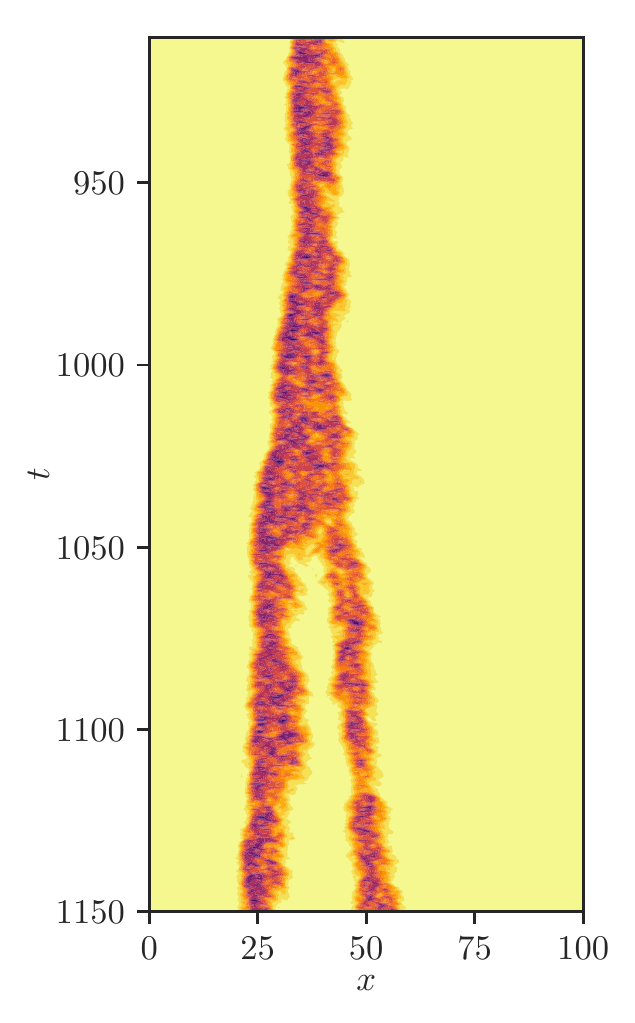}\hspace{-8pt}
    \includegraphics[height=130pt]{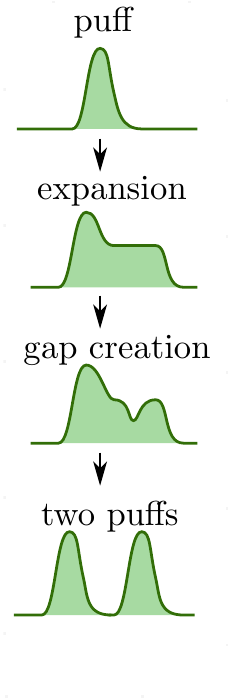}
    \caption{Illustration of puff decay and splitting
      mechanisms. Left: Decay mechanism---Starting from a puff the
      system passes close to the decay edge before turning
      laminar. Splitting mechanism---In the first stage the puff
      expands into a slug with a turbulent core, in the second stage a
      laminar gap is formed within the core. The trajectory passes
      close to the split edge. Trajectories are taken from the
      stochastic Barkley model. \label{fig:example-decay-split}}
  \end{centering}
\end{figure}

The underlying dynamics controlling puff decays are relatively well
understood, as sketched in Fig.~\ref{fig:example-decay-split}
(left). They are driven by rare chaotic fluctuations which push the
system across a phase space boundary between the laminar and puff
state, the so called edge of
chaos~\cite{skufca:2006,lozar:2012,bundar:2019,rolland:2018}. On the
way, the system passes close to a state which lies on this
edge~\cite{schneider:2007,duguet:2008,mellibovsky:2009,avila:2013},
here termed the \textit{decay edge}, whose single unstable direction
mediates the transition.

A comparable dynamical understanding of puff splitting is currently
lacking. The directed percolation picture is predicated on splits
becoming more frequent than decays with increasing $Re$; However, in
the absence of a mechanism for puff splits, it remains unclear if that
is the general rule and under what circumstances this type of
transition could be absent. In this work we propose the first-ever
general mechanism for puff splitting and discuss how it can be
suppressed. The mechanism, sketched in
Fig.~\ref{fig:example-decay-split} (right), is a two stage process: in
the first stage fluctuations drive a puff to expand through a
structure called a slug. Slugs are observed at higher $Re$, where
puffs are absent, and are similar to puffs except for their expanding
core of homogeneous turbulence, see Fig.~\ref{fig:puff-and-slugs}
(right). At the transitional $Re$ we are considering here, we argue
that such structures still exist and would contract as illustrated in
Fig.~\ref{fig:puff-and-slugs} (middle). To expand, they must be driven
by rare fluctuations. In the second stage, when the slug is wide
enough, a laminar pocket forms within the core, separating it into two
parts, each of which evolves its own puff. We suggest that this
transition is mediated by a state we call the \textit{split edge},
lying at the boundary between a one-puff and a two-puff state. In the
following we motivate the viability of this picture for pipes and
other wall bounded flows sharing the same phenomenology. Some of the
ingredients in our mechanism have not been directly observed in these
flows; we explain why we believe they should be present. We validate
the proposed mechanism within the stochastic Barkley
model~\cite{barkley:2015,barkley:2016}, presenting results taken from
simulations, and leave a dedicated study of shear flows by direct
numerical simulations to future work. Finally, we identify how the
proposed split mechanism could be suppressed, discussing possible
signatures.

\section{The slug-gap-split mechanism}
\label{sec:slug-gap-split}

\subsection{Expansion stage} 

\begin{figure}[b]
  \begin{center}
    \includegraphics[width=250pt]{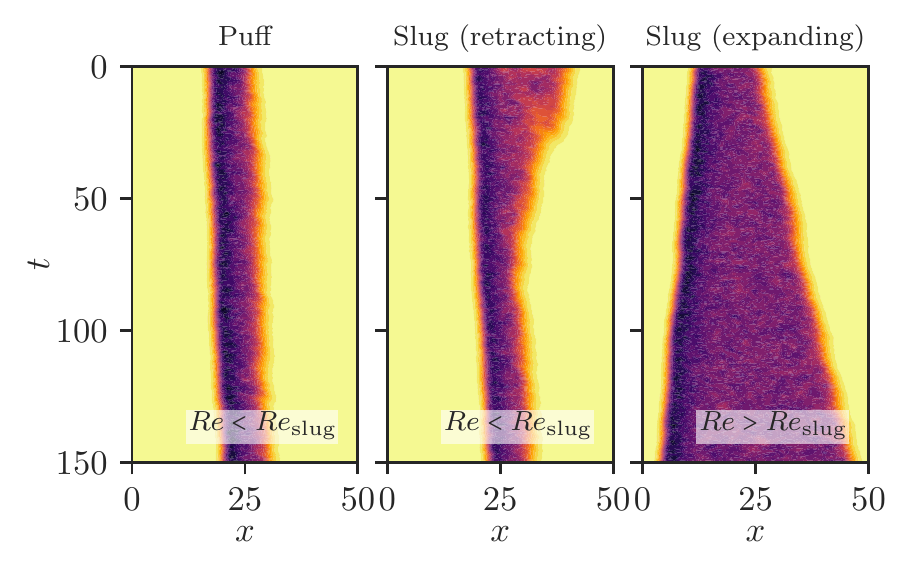}
  \end{center}
  \caption{At $Re<Re_{\text{slug}}$, puffs are stable (left). For
    $Re>Re_{\text{slug}}$, turbulence expands via slugs (right). Slugs
    also exist for $Re<Re_{\text{slug}}$, but they retract into a puff
    instead of expanding (center). Realizations taken from the
    stochastic Barkley model. \label{fig:puff-and-slugs}}
\end{figure}

We begin by motivating a regime of $Re$ where slugs and puffs coexist
for pipe (and duct) flow. A slug is a structure interpolating in space
between a homogeneous turbulent state at its core, where turbulence
production balances turbulence dissipation, and laminar base flow at
its sides. Slugs are observed at $Re>Re_{\text{slug}}$, where they
replace puffs. It is also observed that the expansion rate of a slug
grows continuously with $Re$, starting from zero at
$Re=Re_{\text{slug}}$~\cite{duguet:2010,barkley:2015}. We argue that
this continuous transition implies that homogeneous turbulence can
first be sustained at $Re<Re_{\text{slug}}$ (referred to as a masked
transition in~\cite{barkley:2015}), here denoted by
$Re_{\text{turb}}$, see Fig.~\ref{fig:phase-diagram}. Indeed, that the
relative front speed for slugs continuously increases from zero at
$Re_{\text{slug}}$ implies that the transition from slugs to puffs
with decreasing $Re$ has to do with a change in front speed, rather
than with the disappearance of homogeneous turbulence below
$Re_{\text{slug}}$. If $Re_{\text{slug}}$ was indeed the point where
homogeneous turbulence is first sustained, one would not generically
expect front speeds to match there.

Thus, slugs should be well defined dynamical states also in the range
$Re_{\text{turb}}<Re<Re_{\text{slug}}$, developing when homogeneous
turbulence and laminar flow are brought into spatial contact. The
transition from slugs to puffs with decreasing $Re$ is then a
consequence of the expansion rate of a slug becoming negative in this
range, as is consistent with a continuous decrease in relative front
speed starting from zero at $Re_{\text{slug}}$. We thus expect that,
once excited, a slug would contract and turn into a puff, as is
illustrated in Fig.~\ref{fig:puff-and-slugs} (middle). Such a
contraction of slugs means that laminar flow overtakes homogeneous
turbulence for this range of $Re$, implying that the turbulent state
is metastable~\cite{pomeau:1986}.

Note that we expect contracting slugs to be hard to observe in direct
numerical simulations or experiments: chaotic fluctuations would
quickly split a slug, through the mechanism explained below, if it is
too wide. Indeed, contracting slugs have not been observed in shear
flows to date. They are also hard to observe in the Barkley model for
the classical parameters used in~\cite{barkley:2016}, where the noise
level is much higher than the one we use in
Fig.~\ref{fig:puff-and-slugs}. However, if narrow enough, a
contracting slug should be an observable dynamical state below
$Re_{\text{slug}}$.

For the split mechanism, we propose that the most likely way to expand
a puff is for random fluctuations to overcome the retraction of the
slug. The first stage of our mechanism is thus the expansion of a
puff, via rare chaotic fluctuations, into a slug with a wide enough
turbulent core. This first stage is accessible at
$Re_{\text{turb}}<Re<Re_{\text{slug}}$.

\subsection{Gap formation stage}

The second stage corresponds to the transition from a slug with a
turbulent core to a state with two puffs. For this stage, only the
center of the slug, namely its turbulent core, is relevant. Within
this core, to end up with two separated puffs, a laminar gap must be
formed by chaotic fluctuations.

\begin{figure}[b]
  \begin{center}
    \includegraphics[width=200pt]{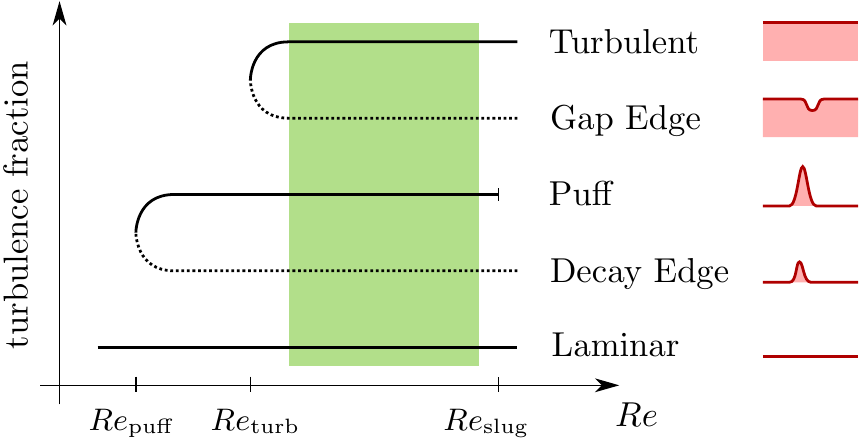}
  \end{center}
  \caption{Bifurcation diagram (sketch) for transitional pipe
    flow. Green shaded region: applicability region of the proposed
    split mechanism. Attracting states are solid lines, unstable edge
    states are dotted. Right column: Sketches of the corresponding
    states. In the deterministic Barkley model used here
    $r_{\text{turb}}=0.667$ and $r_{\text{slug}}=0.726$.
  } \label{fig:phase-diagram}
\end{figure}

When viewed locally, the creation of a laminar gap within turbulent
flow is a transition in its own right. Namely from spatially
homogeneous turbulence everywhere in the pipe into a state with some
laminar flow present. Indeed, homogeneous turbulence should be
metastable for $Re<Re_{\text{slug}}$: a laminar-turbulent front would
overtake the homogeneous turbulent flow, so that an opened laminar gap
would expand and the state would not return to homogeneous
turbulence. Along this transition, there should exist a minimal local
perturbation of the homogeneous turbulence which will open a gap, and
a corresponding mediating edge state: The \textit{gap edge}, see
Fig.~\ref{fig:phase-diagram}.

We expect the gap edge to take the form of a local decrease of
turbulence down to a threshold value. A further decrease of turbulence
would widen the gap until laminar flow is formed, while an increase of
turbulence would close it back. In that sense, the gap edge is the
minimal nucleus of laminarity to create a lasting gap, mirroring the
decay edge between laminar flow and a puff, which is the minimal
nucleus of turbulence to create a turbulent puff.

\begin{figure}[b]
  \begin{center}
    \includegraphics[height=200pt]{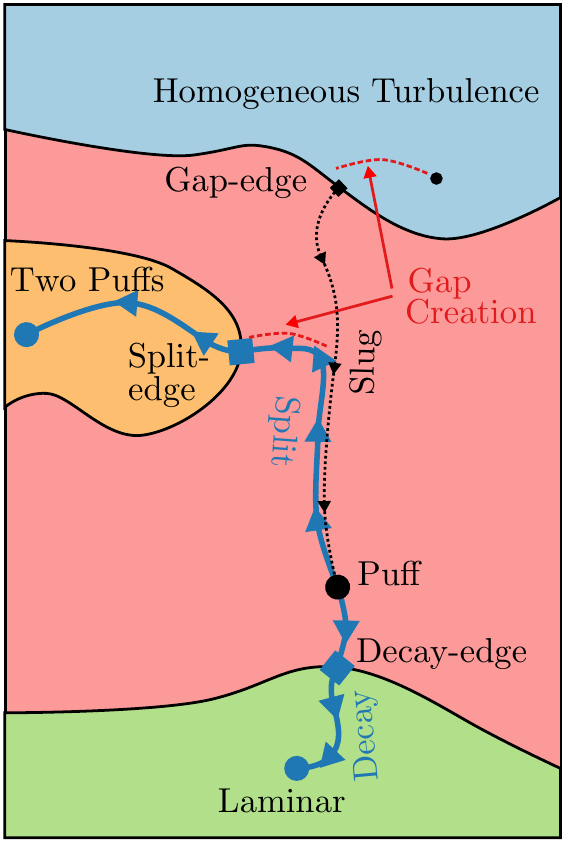}
  \end{center}
  \caption{Sketch of the phase space, transitions and
    edge-states. From the puff basin (red region), the system can
    decay through the decay edge state into the laminar basin (green
    region). Alternatively, a puff can split, transitioning through
    the split edge state into the two puff basin (orange region). The
    split transition (blue line) consists of: \textbf{(i)} a slug
    driven by noise to expand, otherwise deterministically contracting
    (black dotted line), \textbf{(ii)} gap formation within the
    slug. This gap creation (red dotted) is the same as in the
    transition out of the homogeneous turbulence basin (blue
    region). \label{fig:landscape}}
\end{figure}

The terminology \textit{gap edge} implies that we expect this state to
lie on a leaky basin boundary between two long lived states. Indeed,
for periodic boundary conditions, that would be the leaky boundary
between a puff and a homogeneous turbulent state (which can thus be
thought of as the edge of homogeneous chaos). Such a leaky boundary
could form as a result of a boundary crisis, where the homogeneous
turbulence state touches its basin boundary with the puff, making
transitions to the puff state possible. A boundary crisis is also the
mechanism believed to give rise to puff decays. Those are mediated by
the decay edge, passing through the leaky boundary between the laminar
and puff state~\cite{skufca:2006,lozar:2012}. From this point of view,
a contracting slug is a dynamically favorable direction from the
boundary between homogeneous turbulence and a puff, where the gap edge
lies, to the puff see Fig.~\ref{fig:landscape}.

To summarize, we argue that after the expansion into a slug, the
second stage on the way to a puff split is the above described gap
creation, occurring in the turbulent core of the slug.

\subsection{The split edge state}

Viewed as a whole, the split transition requires crossing the boundary
between one and two puffs, motivating yet another edge state---the
\emph{split edge}, see Fig.~\ref{fig:landscape}. Combining the two
stages of the mechanism, the split edge should roughly take the form
of a slug with a gap edge in its core, exactly wide enough to fit it,
as sketched in Fig.~\ref{fig:example-decay-split}. Indeed, local
chaotic fluctuations of the gap edge along its unstable direction can
either widen it, inducing a split, or close the gap, forming the core
of a slug, which would then retract into a puff.

Here, we are suggesting the existence of a leaky boundary between a
single puff and a two-puff state. Transitions from the two puff state
to a one puff state are quite natural: they could occur through a
decay of one of the puffs, implying an edge state of the form of a
puff+decay edge. Such transitions would then be related to a boundary
crisis where the two puff state touches the boundary. However, such an
edge state does not necessarily allow the opposite transition, from a
single puff to two, since the decay edge cannot be spontaneously
excited from laminar flow.  Instead, we expect such transitions to be
mediated by the proposed split-edge state, corresponding to a boundary
crisis where the two-puff state touches the boundary. This might
explain the exponential distribution of transitions times from the
one-puff to the two-puff state observed for puff
splitting~\cite{avila:2011}. The leaky boundary between a one puff and
a two puff state can thus include two embedded edge states, one for
each direction. This complicates splitting events, since after
reaching the two-puff basin, the system can recross the boundary back
to a single puff state, giving rise to near-split events.

\subsection{Role of fluctuations}

In the proposed mechanism, fluctuations have a double role: first,
they drive the puff to become a slug and subsequently expand
it. Second, when the core of the slug is wide enough to facilitate a
gap edge, fluctuations drive the turbulence in the slug core below a
threshold. It then generates an expanding laminar hole, with the two
remaining segments of turbulence naturally evolving into puffs.
Clearly, the expansion stage becomes more likely as $Re$ increases,
slugs contracting increasingly slower as $Re_{\text{slug}}$ is
approached. On the other hand, gap creation becomes less likely with
increasing $Re$, since the homogeneous turbulent state becomes more
stable. The latter trend is opposite to observations in straight
pipes~\cite{avila:2011}, implying that if our mechanism is at play,
the gap creation is not the limiting factor in its splits. Finally
note that the sustainment of the two puff state at the end of the
transition may also depend on fluctuations: the decay probability of
the downstream puff is increased by close proximity to an upstream
puff~\cite{hof:2010,barkley:2016}, causing near-split
events~\cite{gome:2021,SI}.

\section{Results for the Barkley model}
\label{sec:results-barkl-model}

In the remainder, we will demonstrate the relevance of the
slug-gap-split mechanism focusing on transitional turbulence in the
Barkley model. Each step of the analysis we perform could also be
applied to direct numerical simulations of the Navier-Stokes equation,
with appropriate adjustments. However, as it is significantly more
computationally expensive to generate samples for the latter, here we
limit the investigation to the Barkley model. This model is known to
successfully reproduce both qualitative and quantitative features of
pipe and duct flow~\cite{barkley:2015}, relying on minimal modeling
ingredients. Moreover, in the presence of stochastic noise, the model
also goes through a directed percolation transition, facilitated by
puff decays and splits~\cite{barkley:2016}. Note that the
deterministic Barkley model does not exhibit chaos nor proper
turbulence, but rather captures the underlying phase space
structure. In particular, puffs and homogeneous turbulence are
deterministic states in the model. Transitions between basins of
attraction are then made possible by the inclusion of the noise, which
models chaotic fluctuations and allows for leaky boundaries.

The model is one dimensional, describing the coarse grained dynamics
along the pipe direction $x$, and employs two variables: the mean
shear $u(x,t)$ and turbulent velocity fluctuations
$q(x,t)$~\cite{song:2017}. Alternatively, $u(x,t)$ can be interpreted
as the local centerline velocity, which becomes smaller in the
presence of turbulence---namely dropping down to the mean flow rate
$\bar U$, and is largest for laminar flow at $U_0$.

The Barkley model is given by
\begin{equation}
  \label{eq:model}
  \begin{cases}
    \partial_t q +(u-\zeta)\partial_x q=
    f_r(q,u)+D\partial_x^2q+\sigma q \eta \\ \partial_t u +u\partial_x
    u= \epsilon\left[ (U_0-u)+\kappa(\bar{U}-u)q\right]\\
  \end{cases}
\end{equation}
where $f_r(q,u)=q(r+u-U_0-(r+\delta)(q-1)^2)$. The parameter $r$ is
the most important, and plays the role of $Re$. The parameter $D$
controls the strength of turbulence diffusion, $\epsilon$ the (slow)
relaxation of the mean flow to the base laminar profile, $\kappa$ the
influence of turbulence on the mean flow profile (blunting it), and
$\delta$ provides a finite threshold keeping the laminar base flow
stable in the limit $r\to infinity$. Lastly, $\eta$ is a
spatiotemporal white noise with amplitude $\sigma$. It is
multiplicative to mimic the proportionality of chaotic fluctuations in
actual flow to the turbulence level present at that point
(importantly, turbulence cannot be excited from laminar flow where
$q=0$). The values of parameters we choose for our numerical
experiments is discussed in section~\ref{sec:methodology}. The Barkley
model has been demonstrated to quantitatively capture features of
transitional pipe flow remarkably
well~\cite{barkley:2015,barkley:2016}.

In the model, the base laminar flow and homogeneous turbulent state
(as is present in the core of the slug) are fixed points: $q=0, u=U_0$
and $(q_t,u_t)$ correspondingly. Our focus is the range
$r_{\text{turb}}<r<r_{\text{slug}}$ where the turbulent fixed point
coexists with puffs. As sketched in Fig.~\ref{fig:phase-diagram}, in
the model the turbulent fixed point appears in a saddle node
bifurcation at $r=r_{\text{turb}}$ together with an unstable traveling
wave---the gap edge state described above. See~\cite{dynamical_paper}
for a full classification of states.

\begin{figure*}
  \begin{center}
    \includegraphics[width=300pt]{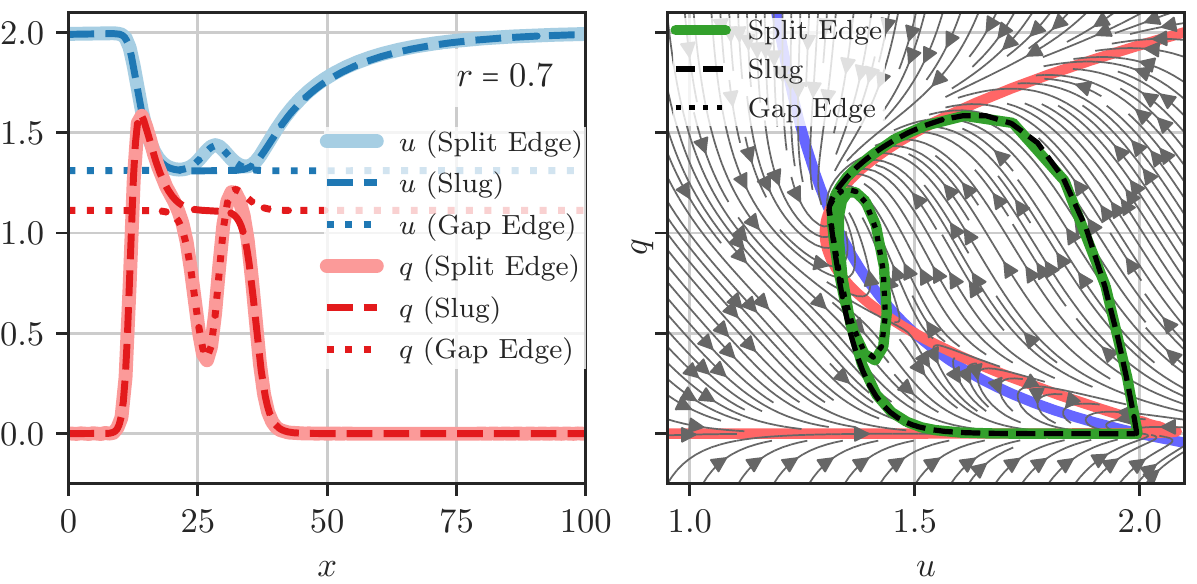}
  \end{center}
  \caption{Deterministic edge state between one and two puffs, the
    \emph{split edge} (solid), overlaid are a slug (dashed), and a gap
    edge (dotted). Configurations are shown in space (left) and in the
    $u$-$q$-plane (right).}
  \label{fig:split-edge}
\end{figure*}

Before turning to stochastic transitions, we analyze the split edge
state lying at the boundary between a one-puff and a two-puff
state. We locate the split edge state using edge tracking within the
deterministic model. Spatial profiles of $q,u$ for the split edge are
shown in Fig.~\ref{fig:split-edge} (left). To confirm the proposed
mechanism is at play, we have also obtained the gap edge for the model
via edge tracking. Superposing the gap edge and a slug with a
(momentarily) equal spatial extent onto the split edge indeed gives an
almost exact match. This match between the three objects is also shown
in a plot in the $q$-$u$-plane in Fig.~\ref{fig:split-edge} (right).

\section{Stochastic transitions in the Barkley model}
\label{sec:stoch-trans}

Decay and split transitions are made possible via random
fluctuations---chaotic in pipe flow and stochastic in the Barkley
model---whose rare realizations bring them about. These transitions
are thus probabilistic in nature, and so must be the comparison to an
underlying dynamical mechanism. We first demonstrate our method of
analysis and the required probabilistic notions for decay transitions,
which are simpler in nature and are already well understood. We then
apply these ideas to test the slug-gap-split mechanism.

\subsection{Puff Decay} 

\begin{figure*}
  \begin{center}
    \includegraphics[height=205pt]{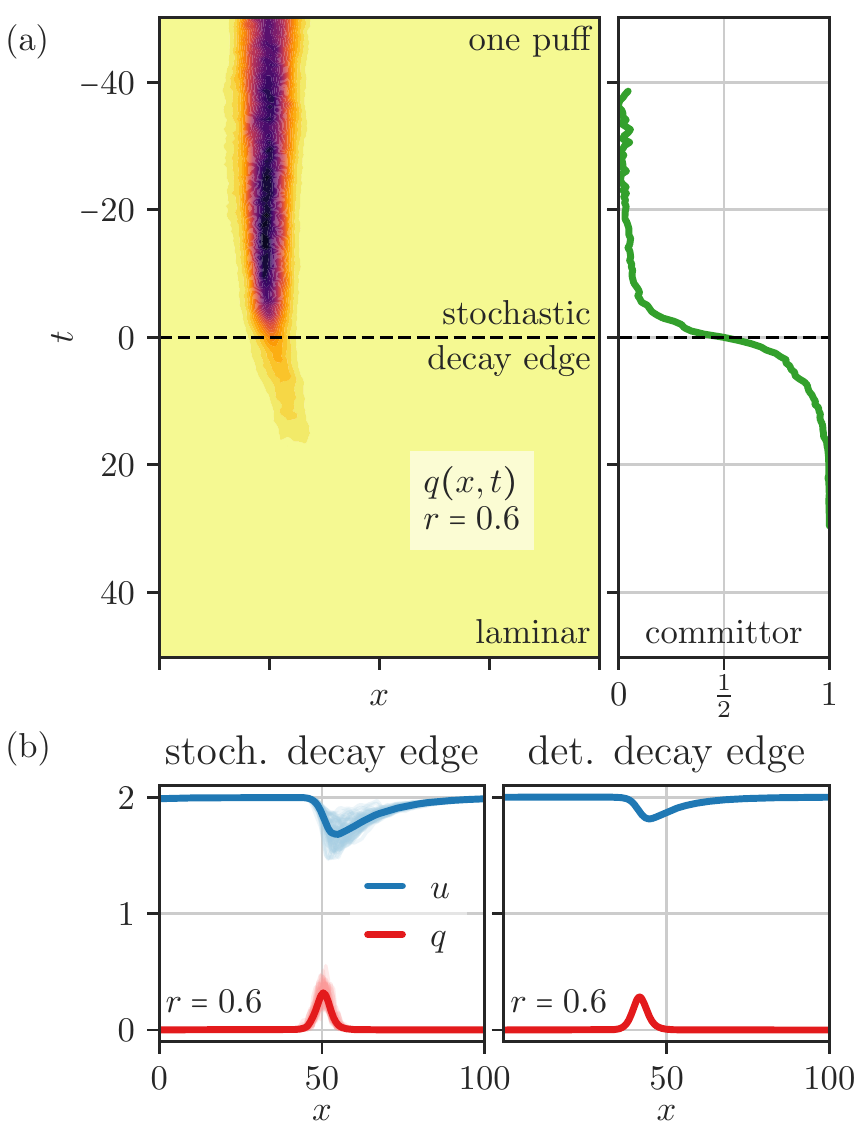}\hspace{-5pt}
    \includegraphics[height=205pt]{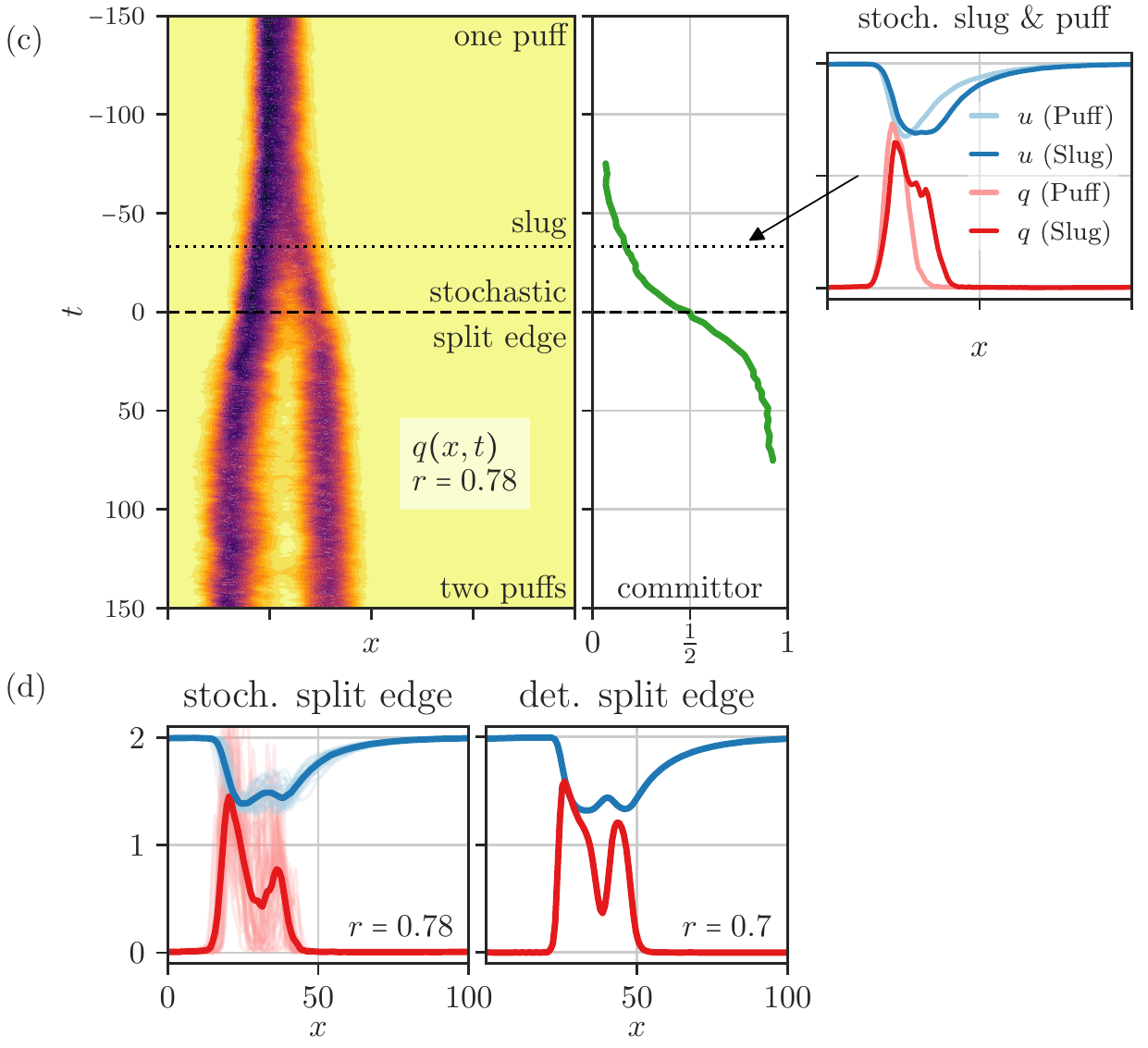}
  \end{center}
  \caption{Transitions in the puff regime,
    $r_{\text{turb}}<r<r_{\text{slug}}$ in the Barkley model where $r$
    is a proxy for $Re$. (a) \textit{Left}: $q(x,t)$ of the average
    stochastic decay trajectory in space-time. \textit{Right}: Average
    committor along the transition. (b) \textit{Left}: Average
    stochastic decay edge, corresponding to the dashed line in (a), on
    top of individual realizations in transparent
    color. \textit{Right}: Deterministic decay edge. (c)
    \textit{Left}: Average stochastic split
    trajectory. \textit{Right}: Average committor. \textit{Inset}:
    Average spatial profile at an initial stage of the transition,
    having a slug-like structure, along with an average puff. (d)
    \textit{Left}: Average stochastic split edge, on top of individual
    realizations. \textit{Right}: Deterministic split
    edge. \label{fig:stochastic-transitions}}
\end{figure*}

We set out to confirm that the decay transition is a trajectory
connecting the puff state to the laminar state, crossing the boundary
through the decay edge. To that aim, we collect many decay
trajectories in the stochastic Barkley model. The average decay path
is shown in Fig.~\ref{fig:stochastic-transitions} as a space-time plot
of the average value of $q$ along the transition. It is presented in
the frame of reference moving at the average speed of a puff. A visual
signature that the trajectory goes through the decay edge is the
increase in speed during the decay, the decay edge moving at a speed
$\approx U_0-\zeta$~\cite{mellibovsky:2009,duguet:2010,barkley:2016}.

In order to speak about an edge between different states in noisy,
stochastic data sets, we further introduce the notion of the
\emph{stochastic decay edge}: This is the set of configurations with
the property of having an equal probability to transition to laminar
flow or to become a puff. Since all observed decay events happen in a
very similar manner, the average of these stochastic decay edge states
is a meaningful state itself. Specifically, we obtain the stochastic
edge in the following way: given an observed decay trajectory we
initiate many stochastic simulations from configurations along it, and
measure the likelihood of continuing on the transition. This defines
the \emph{committor} for the given trajectory~(see
section~\ref{sec:methodology}). We then identify the configuration
from which there is an equal probability to transition or return back,
i.e. where the value of the committor is $1/2$. Repeating this for
many decay trajectories yields the set of configurations defining the
stochastic edge.

We expect the average stochastic edge state to be similar in structure
to the deterministic decay edge. This is confirmed in our numerical
experiments. Fig.~\ref{fig:stochastic-transitions}(b) (left) shows the
spatial profiles of the average stochastic decay edge, averaged over
the different trajectories, on top of individual realizations. For
comparison, a deterministic decay edge is shown in
Fig.~\ref{fig:stochastic-transitions}(b)
(right). In~\ref{fig:stochastic-transitions}(a) we show the average
transition path alongside the committor, averaged over decay
trajectories. Examples of the committor for individual trajectories
are presented in the SI (supplementary information).

\subsection{Puff splitting}

We can now apply these same ideas to the puff splitting
transition. Concretely, to test the relevance of the slug-gap-split
mechanism to stochastic transitions in the Barkley model, we will use
the average transition path obtained from random puff splits. In
Fig.~\ref{fig:stochastic-transitions}(c) we present this average path,
where trajectories are aligned according to the location of the
stochastic split edge. The latter is obtained using the same algorithm
as for the stochastic decay edge; The corresponding average committor
is shown in Fig.~\ref{fig:stochastic-transitions} (c). Note the
striking resemblance of the average transition to a single realization
of a band split in channel flow in a narrow
domain~\cite{gome:2020,gome:2021}. The average transition presented in
Fig.~\ref{fig:stochastic-transitions} (c) indeed reflects the typical
splitting trajectory in the Barkley model, such as that in
Fig.~\ref{fig:example-decay-split}(right); Outliers were included in
the averaging but had a negligible effect. Examples of outlier
splitting trajectories are shown in the SI.

The average transition path clearly shows the expansion stage of the
slug-gap-split mechanism: as seen in the inset of
Fig.~\ref{fig:stochastic-transitions} (c), prior to the creation of
the laminar hole the puff extends into a wider turbulent
structure. Moreover, this average structure resembles a slug,
containing a small homogeneous region most clearly seen in the $u$
profile. Furthermore, the structure always has the same spatial extent
right before gap formation, indicating an expansion stage up to the
necessary length. To test the gap formation stage, we now compare the
average stochastic split edge state to the deterministic split
edge. The stochastic split edge is presented in
Fig.~\ref{fig:stochastic-transitions}(d) (left) on top of individual
realizations, while the deterministic edge is shown
Fig.~\ref{fig:stochastic-transitions}(d) (right). There is very good
qualitative agreement between the two. Note that the stochastic noise
induces parameter shifts, e.g.~of $r_{\text{slug}}$, compared with the
deterministic model~\cite{barkley:2016,SI}; thus we keep the
comparison qualitative.

For the parameters used here, we observe that the main bottleneck for
the transition is the first stage, expansion of the slug. Indeed, once
the slug is wide enough, a laminar gap is likely to form, which smears
the gap within the stochastic split edge in
Fig.~\ref{fig:stochastic-transitions} (d) (left). The likelihood of
splits is thus dominated by the expansion stage, whose likelihood
increases with $r$, as is consistent with observations for the model.

\section{Methodology for the Barkley model analysis}
\label{sec:methodology}

In the following, we will describe the numerical algorithms we
implemented in order to obtain critical points of the dynamics
(i.e.~stable fixed points and edge states), as well the methodology
used to obtain information about the stochastic transition, the
ensemble of transition trajectories, the committor function, and the
stochastic edge state.

\subsection{Stable states}

In order to find the stable states (turbulent, puff, laminar) of the
Barkley model, we numerically integrate it with parameters
$\zeta=0.8$, $\delta=0.1$, $\epsilon = 0.1$, $\kappa=2$, $U_0=2$,
$\bar U=1$, and $D=0.5$, additionally with small diffusion coefficient
for the velocity $D_u=10^{-2}$ which is not expected to affect the
results~\cite{barkley:2016}. The noise amplitude $\sigma$ is
$\sigma=0$ for deterministic simulations and $\sigma=0.5$ for
stochastic simulations, except in Fig.~\ref{fig:puff-and-slugs} where
$\sigma=0.2$ for demonstration purposes. All numerical simulations are
performed in a periodic spatial domain $x\in[0,L]$, with $L=100$.
  
The functions $q(x)$ and $u(x)$ are discretized on an equidistant
computational grid with $N_x=128$ or $N_x=256$ grid points. Spatial
derivatives are computed pseudo-spectrally, by using the fast
Fourier-transform. We use exponential time differencing
(ETD)~\cite{cox-matthews:2002} as temporal integrator, which is exact
for all linear terms, and first order for the nonlinear (reaction)
terms. The time-step is chosen between $\Delta t=10^{-2}$ and $\Delta
t=10^{-3}$. In order to include stochasticity, we generalize
first-order ETD to include the stochastic increment, similar
to~\cite{kloeden-lord-neuenkirch-etal:2011, lord-tambue:2013}.

The puff, gap edge and split edge are traveling at fixed speed along
the pipe, and are therefore not proper fixed points but instead limit
cycles of the dynamics. We fix for that by transforming into a moving
reference frame adaptively, so that the center of turbulent mass of
the objects remains stationary.

\subsection{Edge tracking algorithm}

To find the \emph{unstable} fixed points, i.e. the edge states between
puff and laminar flow (the decay edge), between puff and two puffs
(the split edge), and between turbulent flow and puff (the gap edge),
we employ edge tracking. This algorithm integrates forward in time two
copies of the system, one in each basin of attraction between which
the edge state is to be found. The two copies are kept close via a
bisection procedure to converge back to the separating manifold should
the states drift too far apart. Effectively, this procedure integrates
the system's dynamics, but restricted to the separating manifold
between two stable states. The individual fixed points (laminar, puff,
turbulent, two puffs) are identified via their turbulent mass $\bar
q=\int_0^L q(x)\,dx$.

\subsection{Stochastic transitions}
\subsubsection{Ensemble of Transition Trajectories}
\label{sec:ensemble-trans-traj}

Including stochasticity into the Barkley model, $\sigma\ne0$, allows
the model to \emph{transition} between different meta-stable
states. For example, the puff state is always coexistent with the
laminar state, and fluctuations can drive the puff into eventual
decay.  Numerically, we generate an ensemble of transitions between
two states by initializing in one state, and then simulating the
stochastic dynamics until another state is observed.

For decays, we identify whether a configuration has entered the
laminar state or the turbulent state by checking whether its turbulent
mass $\bar q$ is within an interval of the expected turbulent mass of
the laminar state $\bar q=0$ or the turbulent state $\bar q=q_t$. For
the one puff state, we similarly compare the configuration's turbulent
mass to that of the average puff. It is less straightforward to
identify the two puff state. Here, we flag a potential puff split
event if the fifth Fourier-mode of $q$ exceeds a threshold, which for
our parameters was empirically identified to be sensitive to the
formation of a gap. Whether a split has indeed taken place is then
later checked when computing the committor along the transition and
seeing whether it ever reaches one; see
section~\ref{sec:stoch-edge-track}. This way, we avoid flagging
``near-split'' events, where a turbulent region separates from the
main puff, but is too small to eventually survive.

\subsubsection{Stochastic Edge Tracking}
\label{sec:stoch-edge-track}

In order to compare edge states between the deterministic model (where
edge states can be exactly found, but transitions never happen), and
the stochastic model (where noise-induced transitions can be observed,
but fixed points can only be identified on average), we develop the
notion of the \emph{stochastic edge}. The underlying intuition comes
from the forward committor function known in transition path
theory~\cite{vanden-eijnden:2006, e-vanden-eijnden:2010}: Given a
stochastic process $X_t$ on some state space $\Omega$, consider two
subsets $A\subset\Omega$ the reactant state, and $B\subset\Omega$ the
product state. We are interested in transitions of the process from
$A$ to $B$. The (forward) committor $p^+ : \Omega \mapsto [0,1]$ for
the transition $A\to B$ denotes the probability that the process
visits $B$ next, before visiting $A$. Intuitively, the committor
measures how much the system is ``committed'' to performing the
transition. While the committor can be precisely defined for both
stochastic and deterministic processes~\cite{e-vanden-eijnden:2010},
its computation through solving a Fokker-Planck type equation is
prohobitive for any large system. Instead, for stochastic systems such
as the Barkley model, the committor can be estimated by sampling many
realizations of the process and counting the occurrences of the
transition event.

After generating an ensemble of transitions between two attractors as
described above in section~\ref{sec:ensemble-trans-traj}, we can set
out to numerically compute the committor along these transition
trajectories. One can numerically find the committor of a
configuration $(q,u)$ via sampling: Initialize the simulation at
$(q,u)$ and sample many times, measuring whether we visit the product
state before the reactant state. We do so for many states along each
individual transition trajectory. For example, for a stochastic
transition between the puff and laminar flow, close to the puff the
committor will be almost zero. Close to the laminar state, the
committor will be almost one. In between, there is a region where the
committor takes intermediate values. We define the \emph{stochastic
  edge} of a transition to be a state at which the committor takes the
value $\frac12$. Note that for a stochastic transition a committor
value of $\frac12$ can be attained multiple times per transition. In
our case, we pick as the relevant stochastic edge the state where the
committor is closest to $\frac12$ throughout the transition, and take
the first such state if there are several.

In practice, for the puff decay transition, we define both the set
$A$, the puff, and the set $B$, the laminar state, by thresholding its
turbulent mass $\bar q$. For the puff split transition, we define the
set $A$, the one puff state, and the set $B$, the two puff state, by
thresholding the second cumulant under $q$, i.e.~$\langle x^2\rangle_q
- \langle x\rangle_q^2$. This quantity measures to what degree
turbulent mass is distributed away from the center of turbulent mass,
$\langle x\rangle_q$. It is therefore small for the localized one puff
state, but large for the two puff state, where the center of mass is
located somewhere between the two puffs. The quantity is also large
for extended slugs without a gap which might occur during a transition
event. To avoid mis-identifying a slug for two puffs, we check the
threshold criterion for a prolonged time interval. If the
configuration remains above the threshold very long, then it is almost
guaranteed to be in the (long lived, stationary) two puff state,
instead of the (short lived, transient) extended slug state or other
mixed states, which quickly decay.

\subsubsection{Averaged Stochastic Transition Path}

In order to obtain an average stochastic transition path, we average
all individual trajectories of our ensemble of transition trajectories
obtained as described in~\ref{sec:ensemble-trans-traj}. We average the
committor of individual trajectories over the ensemble of transition
trajectories, by aligning in time the numerically measured committor
at the stochastic edge. Note, though, that the average committor is
not identical to the committor of the average transition
trajectory. We nevertheless show the average committor to give an
impression of how fast the transition happens: The sharpness of the
transition from 0 to 1 indicates the timescale of the transition
itself.

\section{Conclusion and Discussion}

The subcritical transition to turbulence is generically characterized
by the appearance of localized turbulent patches, called puffs in pipe
flow, whose proliferation brings about a sustained turbulent
phase. Here we have presented the first detailed proposal for the
dynamics underlying the proliferation process: a dynamical
puff-splitting mechanism termed the slug-gap-split mechanism. We have
motivated the relevance of this mechanism for pipe flow, and confirmed
its presence in the Barkley model.  The proposed slug-gap-split
mechanism implies concrete predictions, making the proposal
testable. Moreover, it introduces a novel framework within which
previous observations could be interpreted, and alternatives for other
wall-bounded flows could be explored. We now discuss these issues in
detail.

Previously, splitting had been observed to occur through the following
process~\cite{avila:2011,shimizu:2014,mukund:2018}: The puff
continuously emits vortices (turbulence) from its leading edge, then,
if this patch of vortices manages to persists and sufficiently
separate from the parent puff, it seeds a new puff. These observations
could be consistent with the slug-gap-split mechanism, with the growth
of the daughter puff occurring after the crossing of the phase space
boundary between a single puff and two. The possible subsequent decay
would then correspond to a near-split event, a recrossing of this
boundary. Still, such observations could also imply that a different
mechanism is at play, whereby a puff emits a turbulent patch without
going through an expansion stage first. An expansion stage where a
small core of balanced turbulence forms within the puff is thus a
distinct prediction of the slug-gap-split mechanism. Only then does a
laminar gap appear in this picture, and the structure evolves towards
two puffs. The latter process can then sometimes fail if the
downstream puff is snuffed out by the upstream one. It is worth noting
that the splitting process looks much less symmetric in pipe flow
compared to that in the Barkley model. That, however, might be a
visual artifact: the Barkley model does not capture the very steep
increase in turbulence level observed at an upstream front, making the
puff appear more symmetric and thus also the splitting
process\footnote{Indeed, when the turbulence level at the upstream
  front is accentuated for the Barkley model, as done
  in~\cite{barkley:2016}, splits in the Barkley model look very
  similar to those in pipe flow}. This quantitative feature should not
influence the applicability of the slug-gap-split mechanism, which
does not rely on it. In fact, the results presented in
Ref.~\cite{shimizu:2014} for the centerline velocity during a split in
pipe flow do seem to indicate a split through the formation of a
laminar gap within turbulent flow, though further study is needed.
 
To distinguish between the different mechanisms that could be at play
for splits, the analysis outlined above for the Barkley model could be
carried out in direct numerical simulations. First, the split edge
state could be located using edge tracking, as previously done for the
decay edge
state~\cite{schneider:2007,mellibovsky:2009,duguet:2010}. Second, a
probabilistic analysis of split transitions using the committor
between one and two puffs, akin to the one introduced here, could also
be carried out. It would reveal the average transition path and the
stochastic edge state, which could be compared with the expectations
from the slug-gap-split mechanism. Note that the stochastic edge
state, defined here directly via transitional trajectories, is not
a-priori identical to the split edge state found via edge
tracking. The comparison between the two would test the role played by
the latter in the transitions.

The slug-gap-split mechanism is restricted to the range
$Re_{\text{turb}}<Re <Re_{\text{slug}}$. This suggests a novel
possibility that more than a single splitting mechanism exists, and
that different mechanisms could dominate at different $Re$. It would
thus be interesting to assess $Re_{\text{turb}}$ for pipe flow,
e.g. using a minimal flow unit~\cite{jimenez:1991,hamilton:1995}, and
comparing it to the Reynolds number at the directed percolation
critical point. Indeed, while we expect our mechanism to dominate
close to $Re_{\text{slug}}$, it is not guaranteed to survive down to
the critical point. At lower $Re$ it could in principle be replaced by
a process whereby a puff emits a turbulent patch, as described
above. The split edge state would then take a different
character~\cite{SI}.

While we have focused on pipe flow so far, other wall bounded flows
which exhibit a subcritical transition to turbulence (and have a
single extended direction) are captured within the same
framework. Indeed, splits in Couette and channel flow in a narrow
domain seem to follow the proposed mechanism: an expansion stage is
observed, followed by the formation of a laminar
gap~\cite{shi:2013,gome:2020,gome:2021}. Generally, the key condition
for the slug-gap-split to be relevant is for the expansion rate of
turbulence to continuously increase with Re, starting from zero at the
transition from puffs to slugs.

Our work offers a novel point of view on how the phenomenology of
other wall bounded flows could differ from that of pipe flow. In
particular, we now discuss a mechanism by which splits could be
suppressed compared to decays, and therefore a
directed-percolation-type transition would be impossible. This could
be relevant for slightly bent
pipes~\cite{rinaldi:2019,barkley:2019}. In the slug-gap-split
mechanism the likelihood of the transition is the multiplication of
that of the expansion stage, which increases with $Re$, and of the gap
creation stage, which decreases with $Re$. If the latter is
sufficiently high close to $Re_{\text{slug}}$ then transitions would
become more likely with $Re$, as observed in pipes. However, splitting
could be limited by gap creation if such creation becomes improbable
at a sufficiently low $Re$. Then, splits would become less likely with
increasing $Re$ and would be most difficult to observe close to
$Re_{\text{slug}}$. The occurrence of a directed percolation critical
point requires that the probability of puff splitting roughly balances
that of puff decay. Such a balance is not guaranteed if, for large
enough Reynolds numbers, both decays and splits become increasingly
improbable with $Re$. Splits could then remain less probable than
decays for the entire range of $Re$.

A signature that gap creation is indeed a limiting factor for puff
splits would be the absence of an intermittent turbulent regime above
$Re_{\text{slug}}$. This regime, observed in pipe
flow~\cite{moxey:2010}, is characterized by laminar gaps randomly
opening within the homogeneous turbulent state, persisting and
randomly closing. Indeed, if reaching the gap edge is prohibitively
improbable for splits, such laminar gap excitations would also be
suppressed~\cite{dynamical_paper}. In fact, such a correlation seems
to exist for slightly bent pipes~\cite{rinaldi:2019}, providing a
tantalizing connection to the suggested scenario.\vskip6pt

\textbf{Acknowledgments} We thank Dwight Barkley, S\'ebastien Gom\'e
and Laurette Tuckerman for helpful discussions and comments. We are
additionally grateful for input from Yariv Kafri, Dov Levine and
\mbox{Grisha} Falkovich.TG acknowledges support from EPSRC projects
EP/T011866/1 and EP/V013319/1.

\appendix
\section{Noise in the Barkley model}
\label{sec:noise-barkley-model}

As mentioned in the main text, introducing noise changes the
attracting states and their range of existence as compared with the
deterministic dynamics. In particular, front speeds are changed
leading to a change in $r_{\text{slug}}$ so that puffs exist in the
stochastic model in a range of $r$ above that of the deterministic
model, see also~\cite{barkley:2016}.  Evidently, a direct comparison
between deterministic and stochastic states at the same $r$ e.g. for
the the split edge, without properly taking into account the effect of
noise on the deterministic structures, is precluded.

However, the mechanisms presented in the main text do not depend on
such noise effects, even if the realizations of the fluctuations
needed to drive them might. Clarifying the effect of noise on the
invariant solutions of the deterministic model and their regime of
validity, as well as how such changes may affect transition
probabilities and realizations, is left for future work.\\

\section{Example transitions with committor values}
\label{sec:example-trans-with}

\begin{figure*}[p]
  \begin{center}
    \includegraphics[width=240pt]{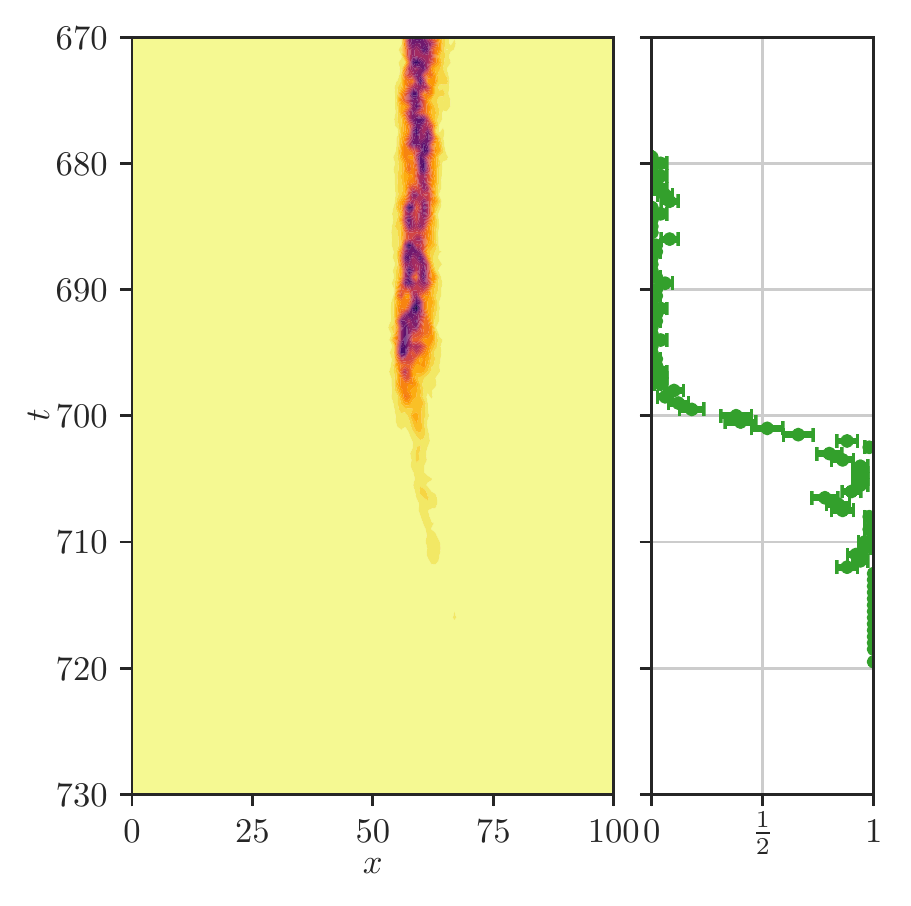}
    \includegraphics[width=240pt]{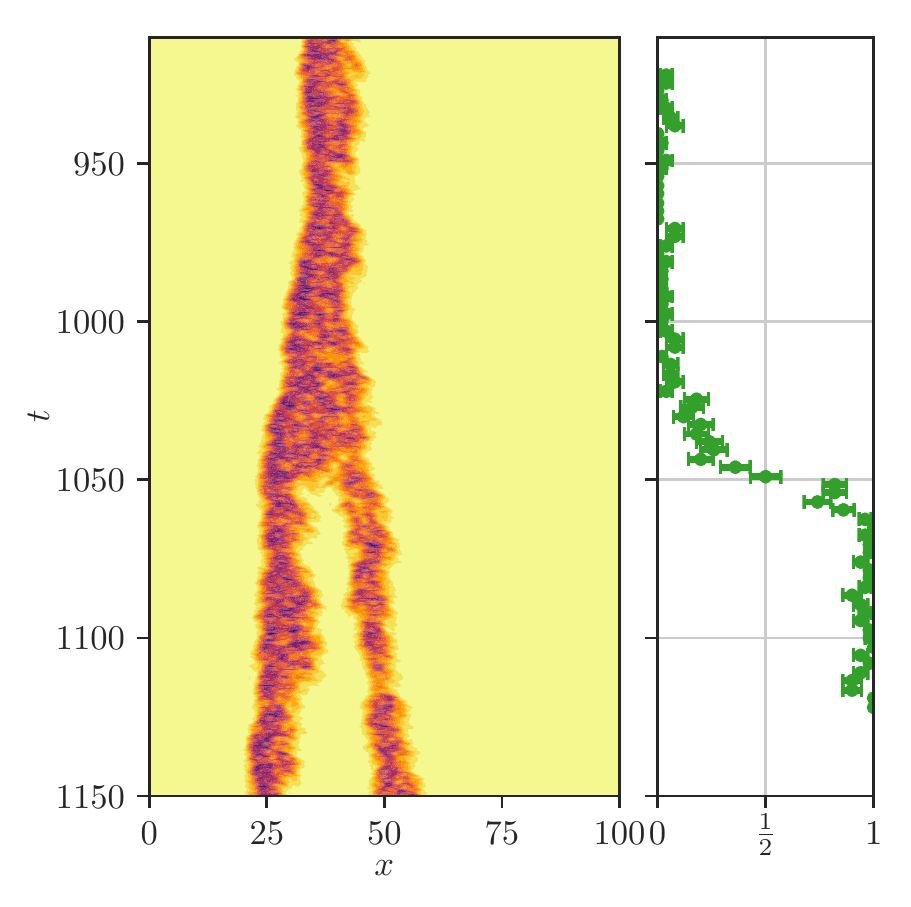}
  \end{center}
  \caption{\emph{Left:} Single stochastic puff decay trajectory,
    together with its committor along the transition. The stochastic
    edge, at committor value $\frac12$, marks the halfway point along
    the transition, where there is an equal probability of continuing
    to decay, or returning back to the puff.\newline\emph{Right:}
    Single stochastic puff split trajectory, together with its
    committor along the transition. The stochastic edge, at committor
    value $\frac12$, is exactly the point at which the puff has
    elongated into a slug, and a gap starts to appear. It demarks the
    point at which returning to a single puff is as likely as
    continuing to split into two.\label{fig:example-decay-split-SI}}
\end{figure*}
\begin{figure*}[p]
  \begin{center}
    \includegraphics[width=240pt]{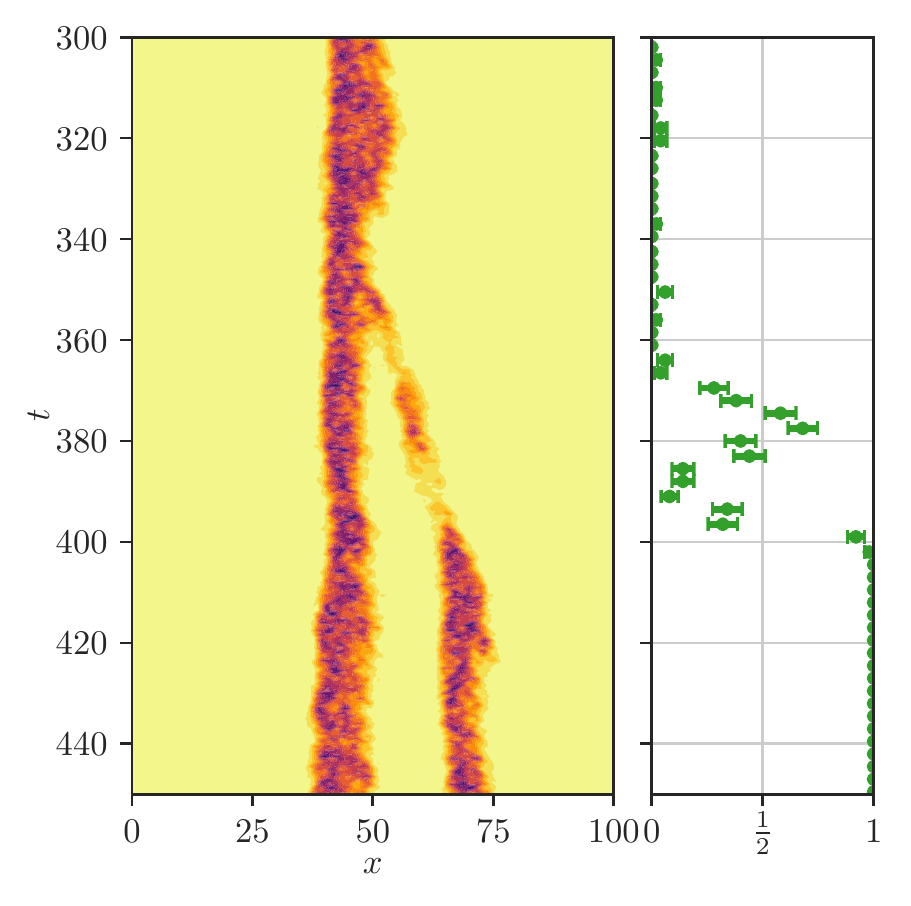}
    \includegraphics[width=240pt]{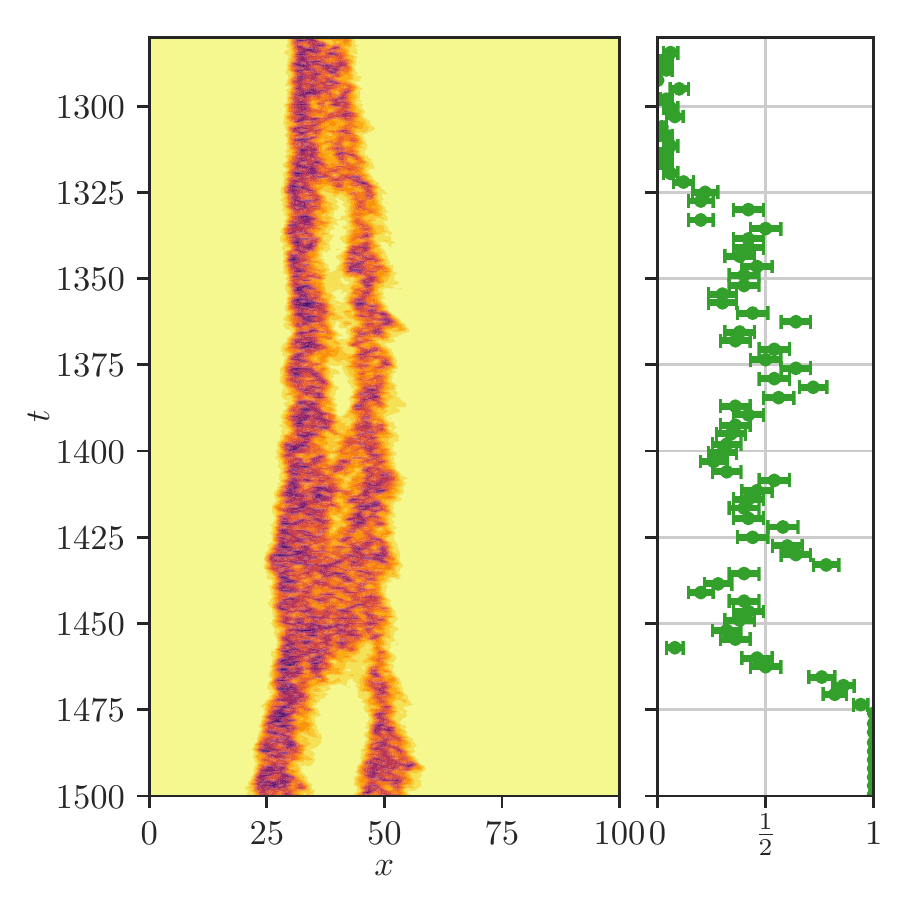}
  \end{center}
  \caption{\emph{Left:}Unusual stochastic puff split trajectory,
    together with its committor along the transition. During the
    initial split, only a very faint puff separates, and is almost
    snuffed out again at $t\approx390$. If it had disappeared, it
    would have been a ``near-split'' event, exhibiting a separating
    puff that is not able to survive. Instead, in this case, a rare
    fluctuation revives the separating puff to full strength, and the
    committor grows again.\newline\emph{Right:} Unusual stochastic
    puff split trajectory, together with its committor along the
    transition. While the initial split looks typical, the trajectory
    re-merges into a very wide slug, skirting the committor-$1/2$
    region for a very long time until eventually separating. Indeed,
    at $t\approx1460$, the secondary puff did almost
    decay.\label{fig:example-nearsplit-averted-ugly}}
\end{figure*}

Figure~\ref{fig:example-decay-split-SI} (left) depicts a single puff
decay trajectory alongside its numerically determined committor
function. For $N_{\text{conf.}}=75$ configurations along the
transition, we took $N_{\text{samples}}=50$ samples to determine the
probability that the process, initialized at that state, would
eventually end up in the puff state or the laminar state. Similarly,
figure~\ref{fig:example-decay-split-SI} (right) shows the same for a
single puff split trajectory. Here it is visible how the committor
takes value $\frac12$ exactly at the point where the gap starts to get
created. Before this point, the process is more likely to return to a
single puff (by retracting the elongated slug state it is in), while
after this point, the process is more likely to continue on with its
split into two puffs by widening the gap between them.

While most split trajectories qualitatively resemble
figure~\ref{fig:example-decay-split-SI} (right), the interaction of
dynamics and noise allow for much more complicated split
trajectories. One class of phenomena are near-split events, where a
secondary puff splits off but is ultimately too small to
survive. Since we condition our split transitions on finally arriving
at two puffs, these near-splits are not included in the ensemble of
split transition trajectories. Yet, its remnants can be felt, as
visible for example in figure~\ref{fig:example-nearsplit-averted-ugly}
(left), where a near-split almost occurs, but is eventually
averted. The single puff splits into two, but the secondary puff is
barely able to survive. As indicated by the committor function, it
would have been far more likely for the secondary puff to decay
again. The chance fluctuation averted the decay, though, and we do end
up with two puffs. In some examples, such as in
figure~\ref{fig:example-nearsplit-averted-ugly} (right), split
transitions meander for an extended time near the
committor-$1/2$-boundary, here, for example, by first splitting, but
re-merging into an unusually wide slug, that eventually splits
again. We stress that for our transition trajectory ensembles we do
not filter out these more complicated events, but instead stick to the
procedure outlined above, and include them in our averages. While this
increases fluctuations around the observed average transition
trajectory and stochastic split edge, it highlights the fact that
these unusual events are comparably unimportant, and are dominated by
the main mechanism that we describe.

\section{Edge state for the transition from two puffs to one}

\begin{figure*}[t]
  \includegraphics[width=.48\textwidth]{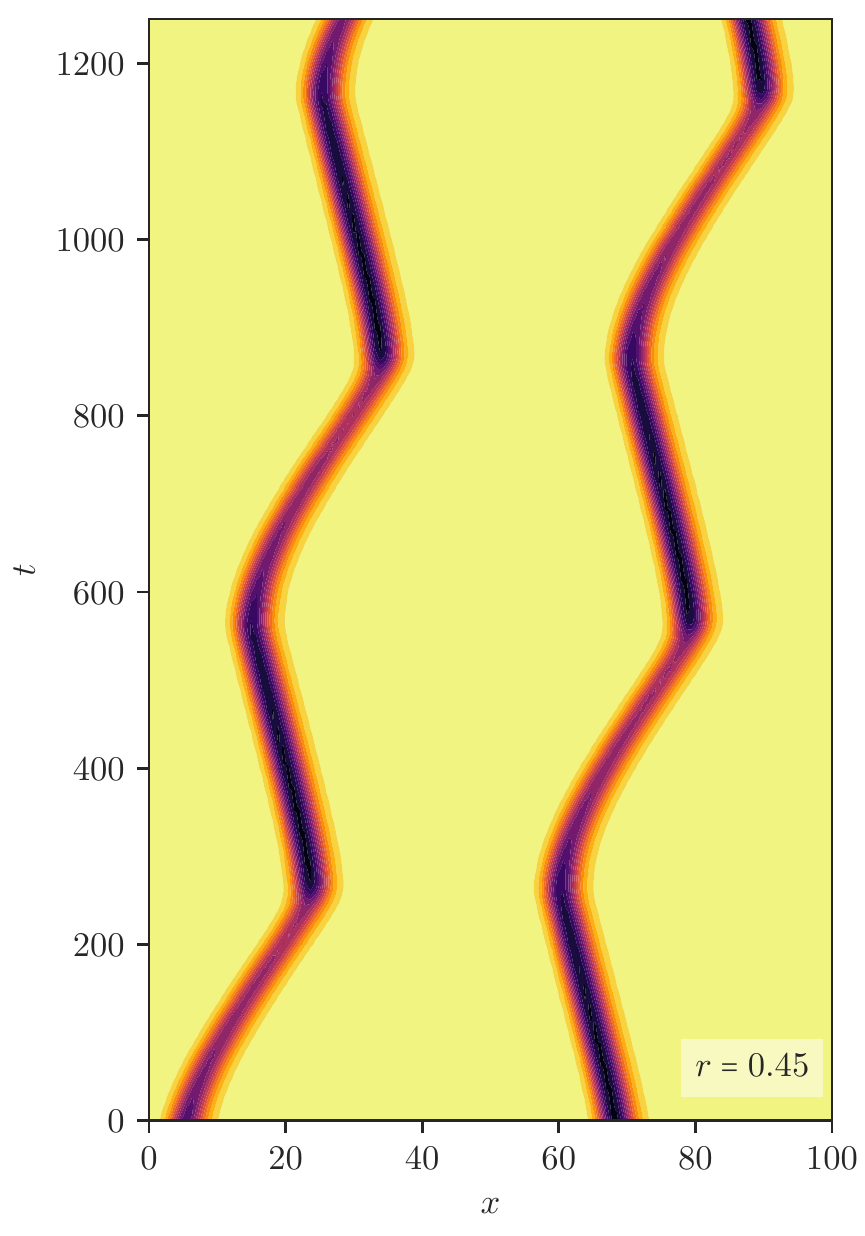}
  \includegraphics[width=.48\textwidth]{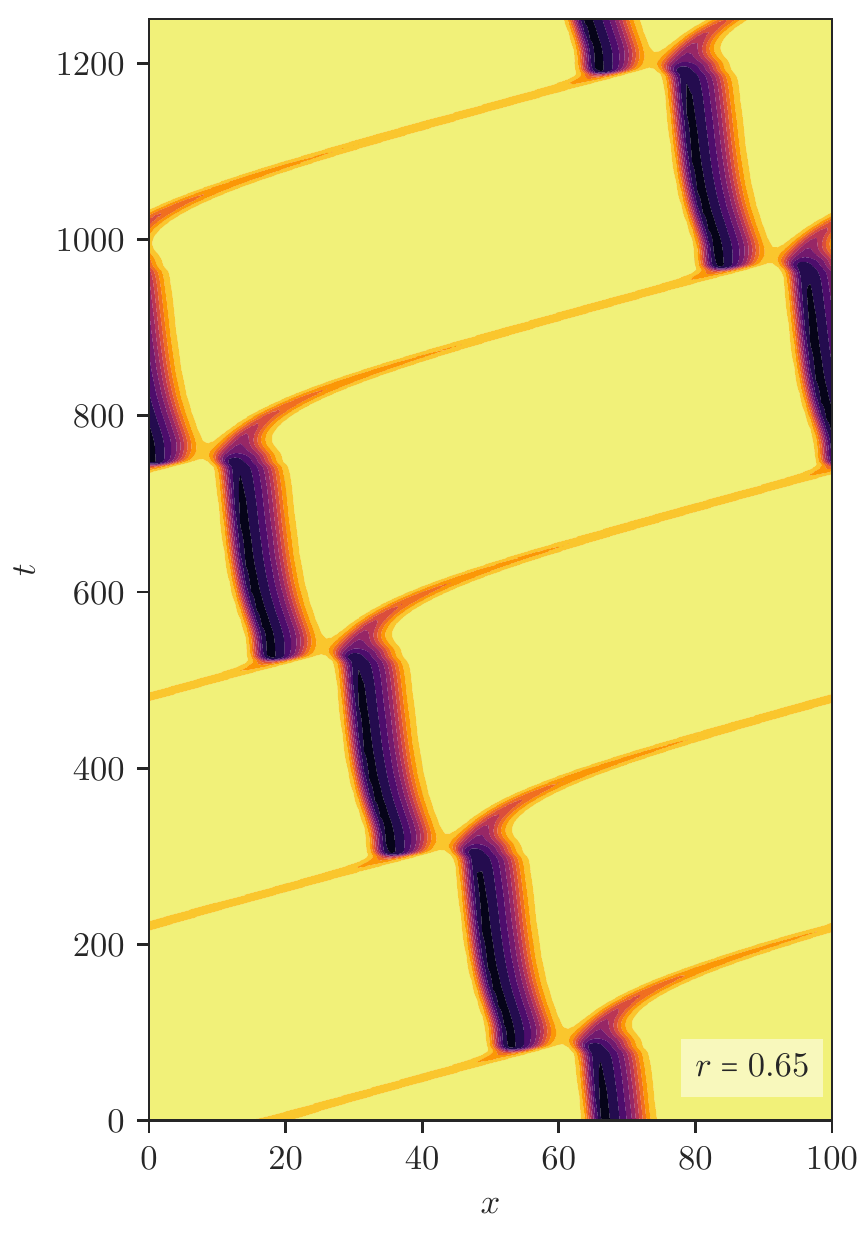}
  \caption{Edge state between the basins of attraction of the one-puff
    state and the two-puff state. Generally, these edge states consist
    of the coexistence of a puff and a decay edge. Since both travel
    at different speed, the resulting (hyperbolic) limiting structure
    cannot be simply a limit cycle in translation. Instead, puff and
    decay edge interact and interchange roles after interaction. This
    is shown for $r=0.45$ (left), where the interaction range is quite
    long, and $r=0.65$ (right), where the interaction range is shorter
    and proper collisions are observed. \label{fig:1-2-puff-edge}}
\end{figure*}

\begin{figure}
  \includegraphics[width=.48\textwidth]{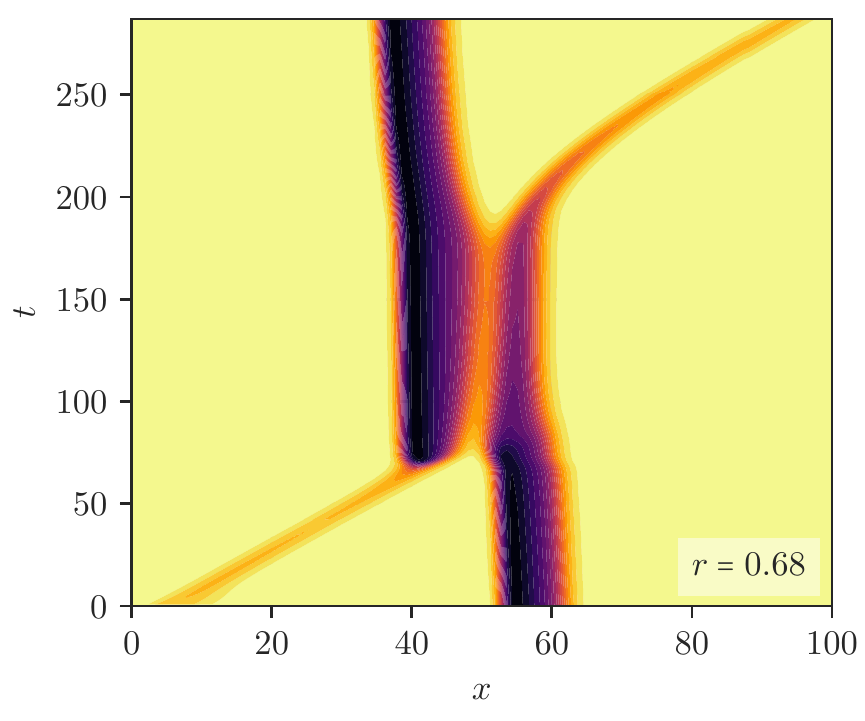}
  \caption{Close-up of the collision portion of the limit cycle on the
    edge between the one- and two-puff states for
    $r=0.68$. \label{fig:1-2-puff-edge-collision}}
\end{figure}

The main material discusses the edge state for a ``puff split'': the
transition between a single puff and a pair of puffs. For the reverse
transition, i.e.~from two puffs to one, a more obvious route would be
the decay of one of the puffs through the single puff decay mechanism,
with the other puff remaining unchanged. An associated edge state
intuitively would be comprised of a puff and a separate decay edge
state far away from it.  In fact, though, such an edge state is quite
complicated. The reason is the following: as highlighted in the main
material, the speed of the decay edge is different from that of the
puff. If we have two puffs present, and want to destroy one of them,
we cannot easily have a stationary state comprised of a puff and a
decay edge, since the two travel at different speeds. In a periodic
domain, they would eventually collide. A fixed point on the border
between a single puff and two puffs can therefore not simply be a
jointly advected combination of puff and decay edge. Instead, the edge
state is a more complicated limit cycle, with co-existence of a puff
and a decay edge at some times, but also collisions between the two
along this limit cycle.

Because of that, the state is no longer a relative fixed
point. Indeed, it is no longer enough to correct for the travel speed
of the edge structure globally, as it consists of multiple, possibly
interacting structures traveling at different speeds. In the projected
coordinate system that eliminates global advection, the edge state
remains a limit cycle. We can still employ the edge tracking
algorithms to restrict the dynamics to the separatrix between the two
basins of attraction, which will consequently evolve into the
complicated periodic structure described above. The result of this can
be seen in figure~\ref{fig:1-2-puff-edge}, where the evolution of $q$
is shown along the hyperbolic edge state, which is indeed periodic (up
to spatial translation). It can be seen that for prolonged periods of
time, this edge state indeed corresponds to a pair of puff and decay
edge, but once their distance becomes too small, they start
interacting, the result of which is an exchange of their roles: The
puff becomes a decay edge, and the decay edge a puff. For $r=0.45$, as
in figure~\ref{fig:1-2-puff-edge} (left), this interaction happens
over considerable distance, while for $r=0.65$ (see
figure~\ref{fig:1-2-puff-edge} (right)), the two collide. Generally,
the two are allowed to approach closer the higher the
$r$. Figure~\ref{fig:1-2-puff-edge-collision} shows a close-up of a
collision between a decay edge and a puff that forms a puff and a
decay edge after the interaction.

Increasing $r$ further, the edge state becomes a single advected
structure, namely the split edge described in the main material. This
configuration is an unstable fixed point and is advected without
deformation. Remnants of the above limit cycle on the separating
manifold still remain, though: The dynamics on the separatrix during
edge tracking quickly converge to a joint puff and decay edge state,
which then slowly evolves until collision, forming the split edge. We
thus hypothesize that the two-to-one-puff transition would never come
close to the split edge state described in the main material, and
instead transition through a state of the form of a puff plus decay
edge, which is no longer an invariant state of the system but a
remnant of the limit cycle at lower $r$. This highlights the
fundamental irreversibility of the system, where forward and backward
transitions are happening with completely different pathways and
mechanisms.

In the Barkley model, the structure of the split edge state thus
changes with $r$ from that where a puff emits a decay edge, as seen in
Figs.\ref{fig:1-2-puff-edge} to the slug+gap structure described in
the main material. The two different edge states thus probably
correspond to two different mechanisms: the slug-gap-split mechanism
which includes an expansion stage and is applicable only above
$r_{\text{turb}}=2/3$, and a mechanism where the puff emits a decay
edge. In the latter, the split transition is expected to happen
through the point in the limit cycle where the two structures are
close by. In the Barkley model, the limit cycle split edge state and
the (relative) fixed point split edge state merge a little above
$r=0.68$, see Fig.~\ref{fig:1-2-puff-edge-collision} where the
collision lasts for a while and visually resembles the slug-plus-gap
edge state. Thus, it appears that the split mechanism whereby a puff
emits a decay edge could be relevant at lower $r$, below or close to
$r_{\text{turb}}$, and is replaced by the slug-gap-split mechanism
once it becomes available.

\section{Comments on the structure of the split edge state in the Barkley model}

In the main text, we have motivated that the split edge should take
the form of a slug with a gap edge in its core. We have also
demonstrated that such a structure is indeed very close to the profile
of a deterministic split edge. However, the actual structure of the
split edge is subtler, since it must be an invariant solution of the
dynamics, in particular a relative fixed point (though there is a
range of $r$ where we observe an edge state in the form of a relative
periodic orbit as described above). This composite object can indeed
move in unison since the speed of the gap edge equals the speed of the
turbulence in the core of a slug. However, if the downstream front was
truly that of a slug, then the split edge would expand. Instead, the
downstream front, which connects directly to the gap edge, must be
closer to that of the puff in structure, the speed of the downstream
front of a puff matching the upstream front and allowing the split
edge to remain stationary in structure. For the value of $r$ shown
(and generally where splits are sufficiently probable to be reasonably
observed) $u_t$ is close to $u_p$ so that it would be hard to
distinguish between the two fronts visually, though some deviations
from a structure of a slug are indeed observed at the point where the
downstream front joins the gap edge, as is expected from the above
argument.

\end{document}